**Digitalization of the IOM: A comprehensive cadaveric study for obtaining three-dimensional models and morphological properties of the forearm's interosseous membrane**

Fabio Carrillo[1,2,*], Simon Suter[1], Fabio A. Casari[1,3], Reto Sutter[4], Ladislav Nagy[1,3], Jess G. Snedeker[2], Philipp Fürnstahl[1]

[1] Research in Orthopedic Computer Science, Balgrist University Hospital, CH-8008 Zurich, Switzerland

[2] Laboratory for Orthopaedic Biomechanics, Institute for Biomechanics, ETH Zürich, CH-8008 Zurich, Switzerland

[3] Department of Orthopaedics, Balgrist University Hospital, CH-8008 Zurich, Switzerland

[4] Radiology, Balgrist University Hospital, CH-8008 Zurich, Switzerland

* **Corresponding Author**:

**Fabio Carrillo, PhD**
Research in Orthopedic Computer Science (ROCS), Balgrist University Hospital
University of Zurich, Forchstrasse 340, CH-8008 Zurich, Switzerland
E-mail: fabio.carrillo@balgrist.ch
Tel. +44 44 510 76 62

**ABSTRACT**

State-of-the-art of preoperative planning for forearm orthopaedic surgeries is currently limited to simple bone procedures. The increasing interest of clinicians for more comprehensive analysis of complex pathologies often requires dynamic models, able to include the soft tissue influence into the preoperative process. Previous studies have shown that the interosseous membrane (IOM) influences forearm motion and stability, but due to the lack of morphological and biomechanical data, existing simulation models of the IOM are either too simple or clinically unreliable. This work aims to address this problematic by generating 3D morphological and tensile properties of the individual IOM structures. First, micro- and standard-CT acquisitions were performed on five fresh-frozen annotated cadaveric forearms for the generation of 3D models of the radius, ulna and each of the individual ligaments of the IOM. Afterwards, novel 3D methods were developed for the measurement of common morphological features, which were validated against established optical ex-vivo measurements. Finally, we investigated the individual tensile properties of each IOM ligament. The generated 3D morphological features can provide the basis for the future development of functional planning simulation of the forearm.

**Introduction**

The musculoskeletal structure of the human forearm enables functions required for the activities of daily living, i.e., pro-/ supination motion, while the surrounding ligament and tendon structures provide stability and allow the translation of forces across the forearm. Post-traumatic bone malunions and soft-tissue injuries can result in a limited range of motion (ROM), cause pain and generate joint instability. The use of clinical static images for the preoperative planning of surgical procedures is the state of the art[1-3]. Previous studies have emphasised the need for dynamic models for the preoperative planning[3-5], in order to study the influence of the surrounding muscles, tendons and ligament on the surgical outcome. However, current simulations used for the analysis of forearm kinematics, rely on bone-only models that are not able to include the influence of the soft tissue into the surgical outcome prediction[1,4-8], and only a few of those forearm models are actually applied in daily clinical practice [7-11]. Existing simulation models are limited by over-simplifications of the biomechanical behaviour, or by the lack of morphological parameters needed for the generation of anatomically accurate models [9,12-14].

Specifically, the kinematics of the pro-supination and the ROM reached by the patient is strongly influenced by the interosseous membrane (IOM), the biggest ligamentous complex in the forearm that connects the radius and ulna bones in the diaphysis [15-22]. Previous studies have demonstrated that the IOM influences bone motion and forearm stability in healthy and pathological cases [18-20,22-26]. Both, in-vivo [22,23] and cadaveric studies [18,20,25] have shown that the motion of the forearm cannot be considered isolated from the IOM and that structures inside the IOM have a major role in the elasticity of the entire ligamentous complex [19,20,26]. In addition, few studies have previously reported the need for a partial release of the IOM in some cases of forearm osteotomy, because the tension exerted by a posttraumatic contracture of the IOM prevented the completion of the planned bone correction [27-29]. From our empirical clinical experience, we have also observed a similar behaviour, in which forearm rotation remained restricted after precise restoration of the geometry of the forearm bones. These findings make clear the need for forearm simulation models capable of including the influence of the soft tissue in order to improve the accuracy of the surgical outcomes [5,6].

For the simulation model of the IOM, several parameters are needed. In the first place, to generate three-dimensional (3D) morphological accurate models of the IOM structures, the thickness, width, ligament attachments and their corresponding insertion angles along the forearm axis are needed. Secondly, for modelling the biomechanical behaviour, the elastic properties of each ligament structure are required, namely, the tensile and shear modulus, the cross-sectional area (CSA), elastic and shear stiffness and maximum tensile and shear forces.

Morphological properties corresponding to each of the individual structures of the IOM have been only reported by Noda, et al. [30] using simple two-dimensional (2D) measurement techniques and without reporting elastic properties. Several cadaveric studies have focused on the description of the biomechanical properties of the IOM [14,18,23,31-36], but most of these studies have only reported on tensile values associated with the entire ligamentous complex [31,33], or have only considered isolated structures, such as the central band [14,34,36] or the distal ligament complex [18,37,38]. Moreover, most of the previous tensile studies have only reported the elastic values associated with shear forces along the longitudinal axis of the forearm [31,33,36]. For the generation of an accurate simulation of the biomechanical behaviour during pro-supination, tensile forces along the ligament fibres are also required. Studies focusing on the analysis of tensile forces along the ligament fibres have only reported on the individual behaviour of the main ligament bundle of the central band [14,34,36], the dorsal and palmar radioulnar ligaments[35,39] or the distal oblique bundle[39-41]. Stabile, et al. [37] performed a comprehensive study about the bi-directional (tensile and shear) mechanical properties of the IOM, however, the reported values were of the entire IOM bundle. In order to generate a simulation capable of considering the individual contributions of the IOM structures on the forearm motion, their corresponding tensile properties are needed.

Thereby, the objective of this study was the generation of 3D morphological parameters and individual tensile properties for each IOM structure of the forearm, namely, the distal oblique bundle, the distal accessory band, the central band and the dorsal oblique accessory cord, based on cadaveric ground-truth data. The following 3D morphological properties were investigated for each ligament of the IOM: radial and ulnar attachment, ligament width, fibre direction, and ligament thickness. In relation to the tensile properties, the CSA, maximum force, stiffnes, and maximum strain were analysed. This work aims to generate ground-truth data for the validation of currently under-development forearm motion models and to provide the basis for the future development of dynamic forearm surgical planning.

**Materials and Methods**

The following procedure was designed for the generation of the 3D morphological and tensile properties of the IOM structures: (1) Exposure of the IOM by removal of muscles and skin, leaving the radius and ulna bones intact; (2) micro-Computer-Tomography (micro-CT) acquisition of the forearms; (3) annotation of the different ligaments structures of the IOM; (4) generation of 3D morphological properties using segmentation and modelling techniques; and finally (5) gathering of the tensile properties of each ligament obtained through tensile test analysis. The evaluation of the obtained 3D morphological properties was done against comparison with established optical ex-vivo measurements.

**Specimen Preparation**

Five fresh-frozen upper extremities cadaver were used in this study. Cadavers were acquired through a donor platform (Science Care, Phoenix, Arizona, USA). Informed consent from next of kin or a legally authorized representative was obtained by the donor platform. The study protocol was approved by the cantonal review board (Zurich cantonal ethical commission, BASEC-Nr. 2018-00282). All forearms were from donors of Caucasian race, with a sex distribution of 3 males and 2 females, and 2 right and 3 left forearms. The mean age of the donors at time of death was 65 years (SD 10.1; range, 50 – 76 years). All forearms were amputated in the middle of the humerus bone. There was no history of previous trauma or pathology. Forearms were stored at -20° C in sealed plastic bags and were thawed overnight at room temperature prior to testing. After removal of the dorsal skin and superficial fascia, all muscles and fat were carefully removed by a senior hand surgeon. The humerus and hand were both carefully dislocated and completely removed from the proximal and distal radioulnar joint. Subsequently, all ligamentous and membranous structures of the IOM were carefully exposed as shown in **Figure 1a.**

**Micro-CT Acquisition**

After preparation, each forearm was packed in a transparent biosafety bag and sealed with a cable-binder. Samples needed to be packed as small and tight as possible due to size limitation of the scanning tunnel of the micro-CT acquisition system; therefore the only possible scanning position for the forearm was in full supination. In order to minimize the signal to noise ratio and possible artefacts during the imaging process, samples had to be completely free of any kind of fiducials. Thus, the annotation process of the ligaments was performed in a later step. Samples were placed into the Micro-CT Scan System (SkyScan 1176, Bruker Kontich, Belgium) and held in place with 3M Micropore surgical tape onto the biggest sample holder available for the Skyscan system. As micro-CT scans have been used in the past mainly for in-vivo and ex-vivo imaging of mice, rabbits and different tissues and material samples, no related work was found corresponding to appropriate device parameters for human forearm samples. After an empirical search of parameters, based on similar micro-CT experiments on animals [42-44], we have designed the following acquisition parameters: slice size 17 µm, energy settings 90 kV and 278 µA, an aluminium filter of 0.5 mm, 210 ms of exposure time, rotation step from 0.8° to 180°, and a frame averaging of 4 frames with a wide field. The approximate scanning speed of the micro-CT was 14 µm /s, meaning that for an averaged-length forearm (~ 250 mm) a complete scan was performed in about 4 hours. The micro-

CT scanning tunnel had a length limitation of 240 mm. Thereby, micro-CT scans were performed from the most distal part of the IOM until the proximal radioulnar joint. For forearms larger than the maximum length (those from male donors), two consecutive scans were acquired: the first scan was done from the distal joint to the 70% most proximal part of the forearm, afterwards, the sample was rotated and a second scan was performed in the same relative position between the radius and ulna (full supination) from the proximal joint to the 70 % most distal part of the forearm, allowing an overlap between both acquisitions.

Afterwards, the acquired micro-CT series were reconstructed using dedicated reconstruction software (NRecon, version: 1.7.4.2, SkyScan, Bruker Kontich, Belgium). An 8-bit reconstruction was done for each series, taking approximately 10 hours and generated a file of 250 GB for a complete dataset. Finally, all reconstructed data were converted into the digital imaging and communication in medicine (DICOM) format by using dedicated commercial conversion software (DicomCT, version 2.5, Bruker, Belgium).

**Annotation of ligaments**

After micro-CT acquisition and following the recommendation of Noda, et al. [30], the different ligamentous structures of the IOM were annotated by a senior hand surgeon. Four different IOM ligament structures were identified across the five forearms (from distal to proximal): the distal oblique bundle (DOB), the distal accessory band (AB), the central band (CB) and the dorsal oblique accessory cord (DOAC) [45]. The AB can be divided into a proximal and a distal part in relation to the CB. In this study, we focused on the AB located distally to the CB. Due to anatomical variations, some structures were missing or were not possible to identify during ligament annotation. The proximal oblique cord (POC) was not recognized in any of the samples. Only 1 forearm exhibited a DOB, which is expected, as only 2 out of 5 individuals exhibit this ligament [18,30]. Radial and ulnar attachments were annotated for each ligament (**Figure 1b-e**) using titanium ligation clips through a clip applicator tool (Ethicon endo-surgery, LLC, USA). Four clips were used for the annotation of DOB (**Figure 1b**), AB (**Figure 1c**) and CB (**Figure 1d**), located at the most proximal and most distal insertion points of the ligament attachment on the radius and on the ulna. Due to its narrow width, the DOAC (**Figure 1e**) was annotated using only two metal clips, located at the centre of its main fibre bundle at the radial and ulnar attachments, respectively. In order to obtain the relative position of the identified ligaments, a clinically standard CT protocol was performed after ligament annotation for each forearm in full supination position, with a slice thickness of 0.6 mm and 120 kV (Somatom Edge Plus, Siemens Medical Systems, Erlangen, Germany). Corresponding DICOM series, for each forearm, were generated using the convolutional kernel (Br 60s/3) for artefact attenuation.

**Generation of 3D morphological properties**

3D models of the IOM were generated from the DICOM files obtained through micro-CT imaging. DICOM files were imported into commercial segmentation software (Mimics Medical, Version 19.0, Materialise 2016, Leuven, Belgium). One complete data set consisted of approximately 10'000 images. Due to RAM, storage, and CPU Power limitation, every data set was divided into subsets of 500 images and an overlap of 100 frames was used. Afterwards, each DICOM subset was loaded into the software and an appropriate threshold for region growing was manually selected. The IOM was manually segmented, slice per slice, to ensure accuracy of the segmentation. Subsequently, radius and ulna bones were segmented in a semi-automatic fashion and corresponding 3D models were created using the Marching Cube algorithm [46] and given in the form of triangular surface meshes (stereolithographic models; hereafter: STL) as described by Roscoe [47].

Similarly, 3D models of the radius and ulna for each forearm were generated from the data obtained through standard-CT acquisition using thresholding and region-growing algorithms of the same segmentation software. 3D models were generated and exported as STL files, together with the surgical clips for 3D identification of the IOM structures.

Consecutively, STL files obtained from the CT and the micro-CT data were imported into the in-house preoperative planning software (CASPA, Balgrist CARD AG, Switzerland). 3D models obtained from micro-CT were combined with the models obtained from standard CT. The positions of all 3D models, including the metal clips, were expressed according to the radius bone model obtained from the CT data. Radius, ulna and IOM model obtained from the micro-CT acquisition were superimposed to the 3D models obtained from the CT, using semi-automatic registration methods[48-51] based on iterative closest point-to-point registration techniques [52] and 3D mesh alignment techniques for medical imaging [53]. In cases were the micro-CT scan had to be performed in two different acquisitions, each of the generated STL models, corresponding to the 70% most distal and 70% most proximal part of the forearms, were aligned to the distal and proximal part of the CT-based 3D models, respectively. Afterwards, a unifying Boolean operation was performed on both STLs to remove overlapping parts. This registration process guaranteed a common reference for all the obtained 3D models and the same position of the ligament attachments along the 3D model of the IOM, generated from the micro-CT.

For each forearm, the relative attachment locations of each ligament were calculated along the radius and ulna, using the position of the corresponding surgical clips. The positions of the clips were also used to calculate the fibre direction vectors for each IOM ligament. For the DOB, AB and CB, two direction vectors were generated at the most proximal (p) and most distal (d) fibres of the ligament as indicated in

**Figure 2**. The direction vector at p was calculated as $\vec{V_p} = U_p - R_p$, where $R_p$ is the proximal insertion point at the radius and $U_p$ is the proximal insertion point at the ulna. Similarly, the direction vector at d was given by $\vec{V_d} = U_d - R_d$. In the case of the DOAC, only one direction vector $\vec{V_m} = U_m - R_m$ was calculated, along the main fibre direction (m in **Figure 2**) indicated by the two metal clips shown in **Figure 1e**.

Subsequently, the IOM model was split into individual 3D ligament models using the polygon clipping algorithm of Vatti [54] and two clipping planes oriented along $\vec{V_p}$ and $\vec{V_d}$ of each ligament. An example of the generation of the 3D model of the CB for one of the forearms is given in **Figure 3**. In the case of the DOAC, the two planes were parallel and oriented according to $\vec{V_m}$ with a width equal to the length of the annotation clips on the ulna and radius.

Four 3D morphological properties were calculated for each of the 5 forearm datasets:

1. The *fan-out angle for each ligament* was calculated from the radial and ulnar attachments, using the direction vectors $\vec{V_p}$ and $\vec{V_d}$. First, the fan-out angle $\theta_p$ was obtained by

$$\theta_p = \cos^{-1}\left(\frac{\vec{V_p} \cdot \vec{A_R}}{|\vec{V_p}||\vec{A_f}|}\right)$$

   where $\vec{A_R}$ is the longitudinal axis of the radius [36]. Subsequently, the fan-out angle $\theta_d$ was calculated at the most distal ligament fibre (d) using $\vec{V_d}$. In the case of the DOAC, only one fan-out angle was calculated based on the centre fibre direction of the main ligament bundle $\vec{V_m}$ (m in **Figure 2**).

2. *Thickness of each ligament*. One advantage of 3D models is the feasibility of measuring the thickness continuously along the entire structure. $T(x_i, y)$ represents the thickness along the ligament fibre $x_i$ of ligament $i$, measured along the cross-sectional plane $y$. We have calculated the thickness $T(x_i, y)$ for each ligament at the most proximal, most distal and the middle point of the ligament as shown in **Figure 2**, with $y$ parallel to $x_i$ (CSA parallel to the ligament fibre directions), with a plane width of 0.2 mm and a resolution step of 0.05 mm. In the case of the DOAC, only the middle ligament fibre was used. An example of the obtained cross-sectional cut for the proximal (p) fibre of the CB is shown in **Figure 4a**. Ligament insertion points on the radius and ulna are indicated in **Figure 4** by the vertical bold grey lines. Subsequently, the thickness of the ligament at each cross-sectional cut was calculated as the maximum vertical distance between the CB points (shown in light pink in **Figure 4a**) along the ligament fibre direction. An example of the calculated $T(x_i = p_{CB}, y \parallel p_{CB})$, thickness of the CB, along fibre direction p, is shown in **Figure 4b.**

3. The *percentage distance of each radial and ulnar attachment* was calculated for each ligament, according to the methodology described by Noda, et al. [30]. First, the 3D radial and ulnar attachment points of each ligament were projected along the corresponding longitudinal axis of each bone, defined according to common anatomical standards [55]. Afterwards, the distance of each ligament attachment was calculated with respect to the most distal points of the ulna and radius, and the attachment location was expressed as the percentage in relation to the total bone length.

4. The radial and ulnar *attachment width* of each ligament were measured as the longitudinal distance between the proximal (p) and distal (d) 3D attachment points of each ligament on the radius and the ulna, respectively.

**Tensile Measurements**

In order to investigate the tensile properties of the IOM structures, each forearm was divided into different ligamentous sections according to their ligament attachments. The ulna of each forearm was carefully resected using a hand saw at the levels of the ligament clips and parallel to the adjacent ligament fibres. The radius was kept intact. Subsequently, each ligament sample was mounted into custom-made bone cups (similar to the setup described by McGinley, et al. [33]) that allowed aligning the fibre direction of the ligament in the direction of the loading force. Radius bone and ulna fragment were fixed to the custom-made bone cups by means of 4-mm diameter screws and reinforced with 2.5-mm diameter Kirschner wires. Afterwards, bone cups were mounted on a universal material testing machine (Zwick 1456, Zwick GmbH, Ulm, Germany). A sketch of the testing setup is shown in **Figure 5**. The pull-out testing was initiated with a preload of 0.5 N and a constant displacement rate of 5 mm/min. A pre-conditioning of 25 cycles was applied to each sample, with 1% strain and 30 mm/min. After the cycling testing, samples were ramp to failure, at a constant displacement rate of 30 mm/min and stopping criteria of 95% of the maximum force.

Applied force, displacement, maximum failure force, and strain were measured for each ligament, as well as failure strains and forces. Data were recorded using TestXpert 10 software (Zwick-Roell, Zwick GmbH, Ulm Germany). The nominal force was calculated based on initial CSA before pull-out. The approximate CSA of each ligament, perpendicular to the loading force, was obtained as the trapezoidal area formed by the average width of the radial and ulnar attachment sites, and the average measured thickness of the ligament along the ligament fibre direction. The elastic mechanical properties were measured in the linear part of the stress-strain curves. First-order polynomials were fitted to the linear range, with the strength for each ligament obtained as the slope of the linear part of the force-displacement curves. Additionally, two peak forces were obtained for each ligament. The first peak happened directly after the linear behaviour of the curve, where the first fibres of the ligaments started to break. The second peak corresponded to the ultimate force, where the rupture of the main ligament fibres was observed. Finally, the ultimate strain of each ligament was obtained from the strain value at which the ultimate force occurred.

**Validation of 3D morphological features**

Ex-vivo measurements of thickness and width were done for each IOM ligament of all forearms. Firstly, ligament fibre directions were marked based on the surgical clips placed on each ligament attachment. Three lines were drawn using a surgical marker: most proximal, most distal and midway between the insertions' sites, corresponding to the ligament fibre directions shown in **Figure 2**. Subsequently, the width of the radial and ulnar attachments of each ligament was measured using a digital micrometre oriented perpendicular to the direction of the fibres.

Afterwards, the CSA of each ligament was measured by averaging the readings of a custom-made linear laser scanner, based on a charge-couple device (CCD; accuracy: $3.52 \pm 1.89\%$, precision: $0.83\%$) [56,57]. The plane of the laser beam was aligned normal to the fibre direction of each sample as shown in **Figure 6a**. The CSA was measured from the radial to the ulnar attachment at the three different levels previously marked (most distal, midway and most proximal fibre of each ligament), and the laser beam was moved along the fibre direction, generating a curve as in the example shown in **Figure 6b** for the DOAC.

The thickness of each ligament was calculated as the distance between the dorsal and palmar surfaces of the cross-sectional curves of **Figure 6b**. Thickness at the radial, ulnar and midway point between the two insertions was recorded for each ligament along the 3 different marked positions (proximal, middle and distal fibres).

## Results

### 3D morphological properties

The obtained 3D models of the 5 forearms, including radius, ulna and the corresponding IOM structures are provided for public download. An example of the segmentation result is depicted in **Figure 7**. The input for the generation of the 3D models was the forearm with exposed IOM (**Figure 7a**). Subsequently, 3D models of the forearm bones and the IOM were generated from micro-CT data (**Figure 7b**). Afterwards, 3D models obtained from micro-CT data were aligned to the CT-based 3D bone models of the radius and ulna (**Figure 7c**). Finally, 3D models of the individual IOM structures for each arm were generated by the described separation method (**Figure 7d**).

The radii had a mean length of 254.4 mm (SD 19.3 mm, range, 230.2 – 274.3 mm) and ulnae 271.5 mm (SD 21.6 mm, range 244.3 – 291.0 mm). Four different 3D morphological properties were analysed for all ligaments: attachment locations, ligament width, ligament angles, and ligament thickness. Obtained values are reported in **Table 1.**

The ligament fibre directions of the CB and AB have their origin at the radius and an ulnar insertion ($\theta$ positive with respect to the longitudinal axis of the radius), in contrast to the fibres of the DOAC and DOB that run in the opposite direction, having their origins at the ulna and their insertion points at the radius ($\theta$ negative with respect to the longitudinal axis of the radius). Subsequently, the ligament widths obtained by the 3D models are compared against the optical measurements in **Table 2**.

In **Figure 8** we compare the thickness measurements obtained from the 3D models against the CCD-Laser measurements along the axial axis and ligament fibre direction. Additionally, we have calculated the average thickness of each ligament, considering both directions, the ligament fibre and the axial axis. A comparison between the values obtained by 3D methods and values recorded from the ex-vivo measurements is given in **Table 3**.

Furthermore, we have analysed the relationship between the thicknesses of the ligaments with respect to their attachment locations along the longitudinal axis of the forearm. Results are shown in **Figure 9** and compared to the findings of McGinley, et al. [58].

**Tensile Properties**

Results of the tensile testing for each IOM ligament of all 5 forearms are given in **Table 4.** The stress-strain curves for each ligament showed a biphasic failure response as described by McGinley, et al. [33] and Stabile, et al. [37]. An initial peak occurred after failure of the thinner fibres of each ligament (first peak force). The second and maximum force peak (ultimate force) happened at the failure point of the thicker and most prominent fibre bundle of the ligament. Linear section of the curves occurred always before the maximum force for all ligaments, between 1% and 3% of the strain for the CB, 1% and 1.5% for the AB, and 1.3% and 2% for the DOAC and the DOB. All specimens of the AB ruptured along the midpoint of the ligament between radius and ulna attachment. In the case of the CB, the structural failure was observed at the proximal ulnar attachment and the distal radial attachment. Specimens of the DOAC broke at the proximal ulnar insertion of the ligament at the maximum load. Similarly, the DOB ruptured along the ulnar attachment of the ligament. Corresponding ultimate failure strains are also reported in **Table 4.**

**Discussion**

The goal of this study was the generation of 3D morphological properties and missing tensile properties of the IOM structures to provide ground-truth data for the validation of the existing forearm motion models. One major advantage of having 3D models of the IOM is the possibility to perform a larger variety of morphological measurements that would be otherwise very time-consuming and sometimes not even possible on ex-vivo measurements. From a clinical perspective, the generated 3D models could help in the construction of realistic forearm simulation based on finite element models, by having a detailed dataset of the morphological properties, the tensile properties, and 3D shape of the IOM. We have made our 3D models public, which represent the first online and publicly available high-resolution 3D models of the IOM.

The obtained 3D morphological features of the IOM structures have been compared against existing anatomical studies and corresponding optical ex-vivo measurements. We discuss the observed differences.

Values obtained from the 3D models of the location of ligament attachments (**Table 1**) and the angle of the fibres of the ligament are in line with the morphological description found in previous anatomical studies [30,36,59]. Nevertheless, 3D attachment locations of the ligaments relative to the bone length, are only in correspondence and within the range for the CB and the DOAC ligaments with respect to the values reported by Noda, et al. [30]. Attachment locations for the AB and the radius insertion of the DOB were located 42% and 39% more proximal, than Noda's reported average values. Similarly, 3D and optically measured widths were on the same range as the values reported by Skahen, et al. [36] for the CB, but two-fold larger for all IOM structures with respect to other anatomical studies [30,34,60]. Noda, et al. [30] used preserved cadaveric forearms, which could incur in shrinkage of the attachment site. Moreover, the lack of consensus in terms of a common anatomical nomenclature of the IOM structures[18,31,34,39,60] could explain the observed differences.

Ligament widths differed by less than 1 mm with respect to the optical ex-vivo measurements (**Table 2**). However, an 8-mm difference was observed between the average width of the optical measurements of ulnar insertion of the CB and the obtained 3D values. This might be associated with the limited accuracy of the calliper measurement method and the difference in forearm positions.

The average thickness calculated by the 3D methods (**Table 3**) differs to those of the ex-vivo measurements in 0.08 mm for the CB, 0.3 mm for the AB and 0.8 mm for the DOAC. The 3D thickness of the DOB was 70% larger than the value obtained by optical ex-vivo measurement. This large difference in thickness is probably due to different forearm positions between CCD-Laser measurements (in neutral position) and micro-CT acquisition (in full supination). The DOB was presumably folded towards the radial end of the ligament, causing the thickness of the fibres to appear larger on the micro-CT acquisition. The observed difference could be also related to an overlap of the DOB with the proximal ligaments of the triangular fibrocartilage complex [30,39]. Additionally, values of the measured thickness by 3D and CCD-Laser were almost two-fold larger than previously reported thickness, done through similar methods [30,34]. Two possible explanations for this difference are (1) different positions of the CSA along the ligament fibre, and (2) different anatomical considerations of fibre bundles. Pfaeffle, et al. [34] did mention possible differences in ligament width and thickness according to the amount of ligamentous tissue considered for the measurement.

A consistent thickness variation was observed for all ligaments along the longitudinal axis of the forearm (**Figure 9**). Thickness at the radial and ulnar attachments are in line with the behaviour reported by McGinley, et al. [58], where the thickness of the ligament decreases gradually towards the proximal end of the forearm. However, ligament thickness measured at the centre of the ligament exhibits an opposite behaviour, with a ligament thickness gradually increasing towards the proximal joint. This behaviour was previously described by Hotchkiss, et al. [31], and it is expected when considering the gradual increase in ligament stiffness among the different ligaments, relative to their location along the longitudinal axis of the forearm (**Table 3**).

The idea of the designed experimental protocol was to obtain information about the tensile properties along the ligament fibres of each IOM ligament. Transversal tensile properties (perpendicular to the ligaments) have been extensively studied in the literature [23,61-63]. Few studies have analysed the tensile stress along the ligament fibre of the CB [34,35,37], but only a reduced set of fibre bundles was considered. Pfaeffle, et al. [34] only reported tensile properties corresponding to the main fibre bundle of the CB, while in our tensile setup the entire CB was considered. The difference in CB fibre bundles affects the CSA and the elastic modulus. Nevertheless, when the difference between CSAs is considered, the elastic modulus scales up to the same order of magnitude of previously reported values. Equivalent tensile values for the remaining ligaments are either inexistent or correspond to values related to shear forces applied perpendicular to the ligament fibre. It is important to note that we do not report on the values of the elastic modulus as we have only considered an uniform and constant CSA for each ligament, taken as the average value among the samples. Nonetheless, the reported stiffness (**Table 4**) is independent of this CSA approximation, as it was directly obtained from the force-displacement curves.

The appearance of a biphasic response in the force-strain curves of our tests confirms the previously reported behaviour of the forearm ligaments under stress testing [31,33,34,64]. The observed rupture point at the middle of the ligament bundles for the AB is expected and consistent with the observed morphology, as it corresponds to the thinner CSA of the ligament. Similarly, the rupture at the radial and ulnar attachments reported for the CB and the DOAC are linked to a smaller CSA at the corresponding attachment points, causing the weaker and thinner ligament fibres of the ligaments to rupture first (thinner CSA), followed by the main fibre bundle. Moreover, out of the 4 ligaments, the CB was the one capable of withstanding a higher load, in average 63% higher than that of the DOB, 46% higher than the AB and 5% higher than the DOAC. This behaviour might be explained by a larger CSA in comparison to the other ligaments. Our results are consistent with previous findings [33,34,36,40,65], where the CB represents the ligament with the higher stiffness among the IOM central and distal structure, providing most likely the most of the stability during pro-supination motion. However, and despite a CSA 74% smaller with respect to that of the CB, the DOAC had a stiffness 77% higher than that of the CB. This behaviour suggests important stabilizing functions of the DOAC on the proximal forearm [34,40]. Having this information can help surgeons to discern between different types of tendon/ligament grafts and surgical techniques for the treatment of IOM injuries. For instance, as the weakest point of the CB ligament is located close to the radial and ulnar attachments, a bone-tendon-bone graft would be recommended over tightrope tenodesis, to avoid possible ruptures[66-68].

This study had the following limitations. First, the small number of samples used allows only for descriptive statistical analysis. Moreover, the acquisition of micro-CT images and subsequently segmentation of the IOM models was a very time-consuming task, limiting its applicability for studies involving a larger sample size. Second, the forearm positions between micro-CT, standard-CT acquisitions, and ex-vivo measurements were not the same. This difference in positions generated small discrepancies between the thickness values obtained by optical measurements and the values obtained from the developed 3D methods. In addition, the width and type of fiducials used for the ex-vivo annotation of the ligaments allow only identifying proximal and distal fibres up to certain accuracy. Also, the 3D morphological properties and corresponding tensile properties of the POC are still missing. The POC was either lost during resection of muscles or could not be identified [30]. Moreover, the calculated CSA was considered to be constant. However, in reality, the ligament width and thickness changes in relation to the applied load. It is also important to mention that some of the cadavers had a smoking history of more than 10 years, which could eventually also affect the reported tensile properties. Finally, our study did not include the proprioception functions of the IOM structures, as the scope of this work focused only on the passive mechanical properties of the IOM structures.

Future studies should focus on the acquisition of the remaining morphological and tensile properties such as the rest length and poison's ratio. Also, the changes in CSA during loading should be analysed in a similar fashion as performed by Scholze, et al. [69]. Furthermore, in order to standardise the relative position between the radius and ulna among the different measurements, we suggest the use of specimen-specific custom-made guides, similar to the currently used 3D-printed guides for intraoperative navigation[2,70-72]. Regarding the anatomical discrepancies with respect to the size of the ligaments' fibres, we would advise performing a standardization of the samples in terms of relative percentage according to the bone size[30]. In this way, anatomical variations among different samples would still be considered, but in a standardized way. In case that the integration of the muscles' influence into the forearm motion is needed, biomechanical experiments would need to be designed to investigate in more detail the forces and torques provided by the individual muscle structures during forearm motion. With the information collected from this study, we plan to build up from a ligament-bone motion model of the forearm and increase the complexity as needed to investigate cases of abnormal tensioning of the IOM structures after forearm bone traumas and osteotomy procedures.

To the best of our knowledge, the present study is the first one to publicly provide 3D models of the ulna, radius and IOM structures, based on cadaveric data. We have successfully generated ground-truth data of the 3D morphological properties of the IOM and provided the basis for validation of existing forearm motion models. Attachment locations, width, and thickness for the IOM structures were provided and validated against established optical ex-vivo measurements and existing literature. Additionally, the tensile properties of the IOM were investigated, including the CSA, strength, maximum force, stiffness and ultimate strain. We believe that the availability of 3D models of the IOM and forearm bones, together with the complete dataset of morphological and biomechanical data, could allow for the improvement of existing forearm simulations. Moreover, we consider that the tensile properties and the 3D shape of the IOM structures could help surgeons to better plan on the most suitable surgical approaches and techniques for the different forearm interventions.

**Data Availability**

The 3D models and datasets generated during the current study are publicly available in our repository: https://rocs.balgrist.ch/fileadmin/user_upload/Forschung/Forearm_3D_Data_download_CarrilloF_2019.zip

**Acknowledgements**

This work was supported by the Swiss National Science Foundation [grant 325230L_163308] and the Balgrist Foundation, Switzerland. Additionally, we would like to thank the radiologists, surgeons and researchers of Balgrist University Hospital for their support and valuable input, especially to Dr. Med. Steven Maurer, Dr. Sander Botter and Elias Bachmann. We would also like to acknowledge the support given by Prof. Daniel Nanz and Natalie Hinterholzer from the Swiss Center for Musculoskeletal Imaging.

**Author Contribution**

F.C. developed the 3D methods and conceived the experimental setup, together with J.G.S. S.S, F.A.C. and L.N prepared the cadaver specimens. F.C, S.S. and F.A.C. conducted the experiments and obtained the morphological data for validation. R.S. developed, supervised and approved the imaging protocol and radiological methodology. F.C, S.S, J.G.S and P.F analysed the results and contributed to the discussion. All pictures and artworks are original work done by F.C. All authors reviewed the manuscript.

**Competing Interest**

The authors declare no competing interests.

# References

1. Fürnstahl, P. *et al.* Automatic and robust forearm segmentation using graph cuts. *Biomedical Imaging: From Nano to Macro, 2008. ISBI 2008. 5th IEEE International Symposium on*, 77-80 (2008).

2. Fürnstahl, P. *et al.* Surgical Treatment of Long-Bone Deformities: 3D Preoperative Planning and Patient-Specific Instrumentation, Computational Radiology for Orthopaedic Interventions. 123-149. (Springer International Publishing, 2016).

3. Jupiter, J. B., Ruder, J. & Roth, D. A. Computer-generated bone models in the planning of osteotomy of multidirectional distal radius malunions. *J Hand Surg Am* **17**, 406-415 (1992).

4. Fohanno, V., Lacouture, P. & Colloud, F. Improvement of upper extremity kinematics estimation using a subject-specific forearm model implemented in a kinematic chain. *J Biomech* **46**, 1053-1059, doi:10.1016/j.jbiomech.2013.01.029 (2013).

5. Moore, D. C. *et al.* Three-dimensional in vivo kinematics of the distal radioulnar joint in malunited distal radius fractures. *J Hand Surg Am* **27**, 233-242 (2002).

6. Hutchinson, D. T., Wang, A. A., Ryssman, D. & Brown, N. A. Both-bone forearm osteotomy for supination contracture: a cadaver model. *J Hand Surg Am* **31**, 968-972, doi:10.1016/j.jhsa.2006.01.010 (2006).

7. Matsuki, K. O. *et al.* In vivo 3D kinematics of normal forearms: analysis of dynamic forearm rotation. *Clin Biomech (Bristol, Avon)* **25**, 979-983, doi:10.1016/j.clinbiomech.2010.07.006 (2010).

8. Weinberg, A. M., Pietsch, I. T., Helm, M. B., Hesselbach, J. & Tscherne, H. A new kinematic model of pro- and supination of the human forearm. *J Biomech* **33**, 487-491, doi:10.1016/s0021-9290(99)00195-5 (2000).

9. Kasten, P., Krefft, M., Hesselbach, J. & Weinberg, A. M. Computer simulation of forearm rotation in angular deformities: a new therapeutic approach. *Injury* **33**, 807-813, doi:10.1016/s0020-1383(02)00114-6 (2002).

10. Kasten, P., Krefft, M., Hesselbach, J. & Weinberg, A. M. Kinematics of the ulna during pronation and supination in a cadaver study: Implications for elbow arthroplasty. *Clin Biomech (Bristol, Avon)* **19**, 31-35, doi:10.1016/j.clinbiomech.2003.08.006 (2004).

11. Kecskeméthy, A. & Weinberg, A. An improved elasto-kinematic model of the human forearm for biofidelic medical diagnosis. *Multibody System Dynamics* **14**, 1-21, doi:10.1007/s11044-005-1756-z (2005).

12. Fürnstahl, P., Schweizer, A., Nagy, L., Szekely, G. & Harders, M. A morphological approach to the simulation of forearm motion. *Conf Proc IEEE Eng Med Biol Soc* **2009**, 7168-7171, doi:10.1109/iembs.2009.5334629 (2009).

13. Péan, F., Carrillo, F., Fürnstahl, P. & Goksel, O. Physical Simulation of the Interosseous Ligaments During Forearm Rotation. *EPiC Series in Health Sciences* **1**, 181-188 (2017).

14. Pfaeffle, H. J. *et al.* A model of stress and strain in the interosseous ligament of the forearm based on fiber network theory. *J Biomech Eng* **128**, 725-732, doi:10.1115/1.2241730 (2006).

15. Hess, F., Farshad, M., Sutter, R., Nagy, L. & Schweizer, A. A novel technique for detecting instability of the distal radioulnar joint in complete triangular fibrocartilage complex lesions. *J Wrist Surg* **1**, 153-158, doi:10.1055/s-0032-1312046 (2012).

16. Ishikawa, J., Iwasaki, N. & Minami, A. Influence of distal radioulnar joint subluxation on restricted forearm rotation after distal radius fracture. *J Hand Surg Am* **30**, 1178-1184, doi:10.1016/j.jhsa.2005.07.008 (2005).

17. King, G. J., McMurtry, R. Y., Rubenstein, J. D. & Ogston, N. G. Computerized tomography of the distal radioulnar joint: correlation with ligamentous pathology in a cadaveric model. *J Hand Surg Am* **11**, 711-717, doi:10.1016/s0363-5023(86)80018-1 (1986).

18. Kitamura, T. *et al.* The biomechanical effect of the distal interosseous membrane on distal radioulnar joint stability: a preliminary anatomic study. *J Hand Surg Am* **36**, 1626-1630, doi:10.1016/j.jhsa.2011.07.016 (2011).

19. Nagy, L., Jankauskas, L. & Dumont, C. E. Correction of forearm malunion guided by the preoperative complaint. *Clin Orthop Relat Res* **466**, 1419-1428, doi:10.1007/s11999-008-0234-3 (2008).

20 Nakamura, T., Yabe, Y., Horiuchi, Y., Seki, T. & Yamazaki, N. Normal kinematics of the interosseous membrane during forearm pronation-supination--a three-dimensional MRI study. *Hand surgery : an international journal devoted to hand and upper limb surgery and related research : journal of the Asia-Pacific Federation of Societies for Surgery of the Hand* **5**, 1-10 (2000).

21 Pfirrmann, C. W. *et al.* What happens to the triangular fibrocartilage complex during pronation and supination of the forearm? Analysis of its morphology and diagnostic assessment with MR arthrography. *Skeletal Radiol* **30**, 677-685, doi:10.1007/s002560100429 (2001).

22 Xu, J. & Tang, J. B. In vivo changes in lengths of the ligaments stabilizing the distal radioulnar joint. *J Hand Surg Am* **34**, 40-45, doi:10.1016/j.jhsa.2008.08.006 (2009).

23 Moritomo, H. *et al.* Interosseous membrane of the forearm: length change of ligaments during forearm rotation. *J Hand Surg Am* **34**, 685-691, doi:10.1016/j.jhsa.2009.01.015 (2009).

24 Pfaeffle, J. *et al.* The stress and strain distribution in the interosseous ligament of the human forearm varies with forearm rotation. Paper presented at Trans 46th Meeting Orthop Res Soc. (2000).

25 Ward, L. D., Ambrose, C. G., Masson, M. V. & Levaro, F. The role of the distal radioulnar ligaments, interosseous membrane, and joint capsule in distal radioulnar joint stability. *J Hand Surg Am* **25**, 341-351, doi:10.1053/jhsu.2000.jhsu25a0341 (2000).

26 Yasutomi, T., Nakatsuchi, Y., Koike, H. & Uchiyama, S. Mechanism of limitation of pronation/supination of the forearm in geometric models of deformities of the forearm bones. *Clin Biomech (Bristol, Avon)* **17**, 456-463, doi:http://dx.doi.org/10.1016/S0268-0033(02)00034-7 (2002).

27 Nagy, L., Jankauskas, L. & Dumont, C. E. Correction of forearm malunion guided by the preoperative complaint. *Clinical orthopaedics and related research* **466**, 1419-1428, doi:10.1007/s11999-008-0234-3 (2008).

28 Sarmiento, A., Ebramzadeh, E., Brys, D. & Tarr, R. Angular deformities and forearm function. *Journal of Orthopaedic Research* **10**, 121-133, doi:10.1002/jor.1100100115 (1992).

29 Matthews, L. S., Kaufer, H., Garver, D. F. & Sonstegard, D. A. The effect on supination-pronation of angular malalignment of fractures of both bones of the forearm. *The Journal of bone and joint surgery. American volume* **64**, 14-17 (1982).

30 Noda, K. *et al.* Interosseous membrane of the forearm: an anatomical study of ligament attachment locations. *J Hand Surg Am* **34**, 415-422 (2009).

31 Hotchkiss, R. N., An, K. N., Sowa, D. T., Basta, S. & Weiland, A. J. An anatomic and mechanical study of the interosseous membrane of the forearm: pathomechanics of proximal migration of the radius. *J Hand Surg Am* **14**, 256-261 (1989).

32 Malone, P. S., Cooley, J., Morris, J., Terenghi, G. & Lees, V. C. The biomechanical and functional relationships of the proximal radioulnar joint, distal radioulnar joint, and interosseous ligament. *J Hand Surg Eur Vol* **40**, 485-493, doi:10.1177/1753193414532807 (2015).

33 McGinley, J. C., D'Addessi, L., Sadeghipour, K. & Kozin, S. H. Mechanics of the antebrachial interosseous membrane: Response to shearing forces. *J Hand Surg Am* **26**, 733-741, doi:https://doi.org/10.1053/jhsu.2001.24961 (2001).

34 Pfaeffle, H. J. *et al.* Tensile properties of the interosseous membrane of the human forearm. *J Orthop Res* **14**, 842-845, doi:10.1002/jor.1100140525 (1996).

35 Schuind, F. *et al.* The distal radioulnar ligaments: A biomechanical study. *The Journal of Hand Surgery* **16**, 1106-1114, doi:http://dx.doi.org/10.1016/S0363-5023(10)80075-9 (1991).

36 Skahen, J. R., 3rd, Palmer, A. K., Werner, F. W. & Fortino, M. D. The interosseous membrane of the forearm: anatomy and function. *J Hand Surg Am* **22**, 981-985, doi:10.1016/s0363-5023(97)80036-6 (1997).

37 Stabile, K. J., Pfaeffle, J., Weiss, J. A., Fischer, K. & Tomaino, M. M. Bi-directional mechanical properties of the human forearm interosseous ligament. *J Orthop Res* **22**, 607-612, doi:10.1016/j.orthres.2003.05.002 (2004).

38  Sun, J. *et al.* Construction and Validation of a Three-Dimensional Finite Element Model of the Distal Radioulnar Joint. *Journal of Mechanics in Medicine and Biology* **16**, 1650010 (2016).

39  Okada, K. *et al.* Morphological evaluation of the distal interosseous membrane using ultrasound. *Eur J Orthop Surg Traumatol* **24**, 1095-1100, doi:10.1007/s00590-013-1388-6 (2014).

40  Stabile, K. J., Pfaeffle, J., Saris, I., Li, Z. M. & Tomaino, M. M. Structural properties of reconstruction constructs for the interosseous ligament of the forearm. *J Hand Surg Am* **30**, 312-318, doi:10.1016/j.jhsa.2004.11.018 (2005).

41  Watanabe, H., Berger, R. A., Berglund, L. J., Zobitz, M. E. & An, K. N. Contribution of the interosseous membrane to distal radioulnar joint constraint. *J Hand Surg Am* **30**, 1164-1171, doi:10.1016/j.jhsa.2005.06.013 (2005).

42  De Schaepdrijver, L., Delille, P., Geys, H., Boehringer-Shahidi, C. & Vanhove, C. In vivo longitudinal micro-CT study of bent long limb bones in rat offspring. *Reproductive Toxicology* **46**, 91-97, doi:10.1016/j.reprotox.2014.03.004 (2014).

43  Saito, S. & Murase, K. Detection and Early Phase Assessment of Radiation-Induced Lung Injury in Mice Using Micro-CT. *PLoS One* **7**, e45960, doi:10.1371/journal.pone.0045960 (2012).

44  Voor, M. J., Yang, S., Burden, R. L. & Waddell, S. W. In vivo micro-CT scanning of a rabbit distal femur: repeatability and reproducibility. *J Biomech* **41**, 186-193, doi:10.1016/j.jbiomech.2007.06.028 (2008).

45  Rein, S. *et al.* Immunofluorescence analysis of sensory nerve endings in the interosseous membrane of the forearm. *Journal of Anatomy* (2019).

46  Lorensen, W. E. & Cline, H. E. Marching cubes: A high resolution 3D surface construction algorithm. Paper presented at ACM siggraph computer graphics. ACM. (1987).

47  Roscoe, L. Stereolithography interface specification. *America-3D Systems Inc* **27** (1988).

48  Kawakami, H. *et al.* 3D analysis of the alignment of the lower extremity in high tibial osteotomy. Paper presented at International Conference on Medical Image Computing and Computer-Assisted Intervention. Springer. (2002).

49  Schenk, P., Vlachopoulos, L., Hingsammer, A., Fucentese, S. F. & Fürnstahl, P. Is the contralateral tibia a reliable template for reconstruction: a three-dimensional anatomy cadaveric study. *Knee Surg Sports Traumatol Arthrosc* **26**, 2324-2331, doi:10.1007/s00167-016-4378-5 (2018).

50  Schweizer, A., Fürnstahl, P., Harders, M., Szekely, G. & Nagy, L. Complex radius shaft malunion: osteotomy with computer-assisted planning. *Hand (N Y)* **5**, 171-178, doi:10.1007/s11552-009-9233-4 (2010).

51  Vlachopoulos, L., Carrillo, F., Gerber, C., Szekely, G. & Fürnstahl, P. A Novel Registration-Based Approach for 3D Assessment of Posttraumatic Distal Humeral Deformities. *J Bone Joint Surg Am* **99**, e127, doi:10.2106/jbjs.16.01166 (2017).

52  Besl, P. J. & McKay, N. D. Method for registration of 3-D shapes. *Sensor Fusion IV: Control Paradigms and Data Structures* **1611**, 586-607 (1992).

53  Audette, M. A., Ferrie, F. P. & Peters, T. M. An algorithmic overview of surface registration techniques for medical imaging. *Med Image Anal* **4**, 201-217, doi:10.1016/s1361-8415(00)00014-1 (2000).

54  Vatti, B. R. A generic solution to polygon clipping. *Communications of the ACM* **35**, 56-63 (1992).

55  Wu, G. *et al.* ISB recommendation on definitions of joint coordinate systems of various joints for the reporting of human joint motion—Part II: shoulder, elbow, wrist and hand. *Journal of biomechanics* **38**, 981-992 (2005).

56  Fessel, G., Cadby, J., Wunderli, S., van Weeren, R. & Snedeker, J. G. Dose- and time-dependent effects of genipin crosslinking on cell viability and tissue mechanics - toward clinical application for tendon repair. *Acta Biomater* **10**, 1897-1906, doi:10.1016/j.actbio.2013.12.048 (2014).

57  Vergari, C. *et al.* A linear laser scanner to measure cross-sectional shape and area of biological specimens during mechanical testing. *J Biomech Eng* **132**, 105001, doi:10.1115/1.4002374 (2010).

58  McGinley, J. C., Roach, N., Gaughan, J. P. & Kozin, S. H. Forearm interosseous membrane imaging and anatomy. *Skeletal Radiol* **33**, 561-568, doi:10.1007/s00256-004-0795-5 (2004).

59  Farr, L. D., Werner, F. W., McGrattan, M. L., Zwerling, S. R. & Harley, B. J. Anatomy and Biomechanics of the Forearm Interosseous Membrane. *J Hand Surg Am* **40**, 1145-1151.e1142, doi:10.1016/j.jhsa.2014.12.025 (2015).

60  Nakamura, T., Yabe, Y., Horiuchi, Y. & Yamazaki, N. Three-dimensional magnetic resonance imaging of the interosseous membrane of forearm: a new method using fuzzy reasoning. *Magn Reson Imaging* **17**, 463-470, doi:https://doi.org/10.1016/S0730-725X(98)00183-0 (1999).

61  Birkbeck, D. P., Failla, J. M., Hoshaw, S. J., Fyhrie, D. P. & Schaffler, M. The interosseous membrane affects load distribution in the forearm. *The Journal of Hand Surgery* **22**, 975-980, doi:https://doi.org/10.1016/S0363-5023(97)80035-4 (1997).

62  Hotchkiss, R. N., An, K.-N., Sowa, D. T., Basta, S. & Weiland, A. J. An anatomic and mechanical study of the interosseous membrane of the forearm: Pathomechanics of proximal migration of the radius. *The Journal of Hand Surgery* **14**, 256-261, doi:http://dx.doi.org/10.1016/0363-5023(89)90017-8 (1989).

63  Pfaeffle, H. J. *et al.* Tensile properties of the interosseous membrane of the human forearm. *Journal of Orthopaedic Research* **14**, 842-845, doi:10.1002/jor.1100140525 (1996).

64  Gabriel, M. T., Pfaeffle, H. J., Stabile, K. J., Tomaino, M. M. & Fischer, K. J. Passive strain distribution in the interosseous ligament of the forearm: implications for injury reconstruction. *J Hand Surg Am* **29**, 293-298, doi:10.1016/j.jhsa.2003.10.023 (2004).

65  Wallace, A. L., Walsh, W. R., van Rooijen, M., Hughes, J. S. & Sonnabend, D. H. The interosseous membrane in radio-ulnar dissociation. *J Bone Joint Surg Br* **79**, 422-427 (1997).

66  Gaspar, M. P. *et al.* Late Reconstruction of the Interosseous Membrane with Bone-Patellar Tendon-Bone Graft for Chronic Essex-Lopresti Injuries: Outcomes with a Mean Follow-up of Over 10 Years. *J Bone Joint Surg Am* **100**, 416-427, doi:10.2106/jbjs.17.00820 (2018).

67  Meals, C. G., Forthman, C. L. & Segalman, K. A. Suture-Button Reconstruction of the Interosseous Membrane. *J Wrist Surg* **5**, 179-183, doi:10.1055/s-0036-1584547 (2016).

68  Tejwani, S. G., Markolf, K. L. & Benhaim, P. Reconstruction of the interosseous membrane of the forearm with a graft substitute: a cadaveric study. *J Hand Surg Am* **30**, 326-334, doi:10.1016/j.jhsa.2004.05.017 (2005).

69  Scholze, M. *et al.* Utilization of 3D printing technology to facilitate and standardize soft tissue testing. *Scientific Reports* **8**, 11340, doi:10.1038/s41598-018-29583-4 (2018).

70  Dobbe, J. G. G. *et al.* Computer-Assisted Planning and Navigation for Corrective Distal Radius Osteotomy, Based on Pre- and Intraoperative Imaging. *Biomedical Engineering, IEEE Transactions on* **58**, 182-190, doi:10.1109/TBME.2010.2084576 (2011).

71  Miyake, J. *et al.* Three-Dimensional Corrective Osteotomy for Malunited Diaphyseal Forearm Fractures Using Custom-Made Surgical Guides Based on Computer Simulation. *JBJS Essential Surgical Techniques* **2**, e24, doi:10.2106/jbjs.st.l.00022 (2012).

72  Rosseels, W., Herteleer, M., Sermon, A., Nijs, S. & Hoekstra, H. Corrective osteotomies using patient-specific 3D-printed guides: a critical appraisal. *European Journal of Trauma and Emergency Surgery*, doi:10.1007/s00068-018-0903-1 (2018).

**Figure Legends**

**Figure 1.** IOM exposure and annotation ligaments. (a) Forearm after exposure. IOM is indicated by the white arrow. (b-e) Annotation of IOM structures with surgical clips. (b) DOB (distal joint to the left); (c) AB (distal joint to the left); (d) CB (distal joint to the right) and (e) DOAC (distal joint to the right). Locations of the surgical clips are indicated in red.

**Figure 2.** Ligament fibre directions. Three fibre directions (shown with dotted arrows) were defined at the proximal "p", middle "m" and distal "d" part of each the ligament. Axial axis of the forearm is shown in red.

**Figure 3.** Generation of 3D model of the Central Band of the IOM. The segmented IOM obtained from the micro-CT is shown in light grey and separate central band in pink. 3D attachment points on the radius and ulna, including the middle points, are indicated with purple arrows and depicted on the membrane as purple spheres. Clipping planes used for the generation of the 3D model of the CB are shown in grey.

**Figure 4.** Thickness Measurement for the 3D models of the IOM structures. (a) Cross-sectional cut of the radius, ulna and IOM surface meshes, generated by a plane aligned at the proximal fibre (p) of the central band (CB) ligament of the IOM. (b) Thickness values for the CB calculated from the cross-sectional cut along the proximal fibre (p).

**Figure 5.** 2D Sketch of the measurement setup used for the tensile experiments. Adjustable positions and angles are indicated by the red arrows.

**Figure 6.** CSA Measurement of ligaments. (a) Example setup of the CCD-Laser, with the laser beam aligned normal to the direction of the ligament fibre. (b) Example of CSA measurement obtained for the DOAC of one of the forearms. The multiple lines represent the three measurements done for the dorsal and palmar surfaces of the ligament.

**Figure 7.** Generated 3D models for forearm 1. (a) IOM sample corresponding to forearm 1, shown after resection of all muscles and before the corresponding marking of the different ligament structures. (b) Obtained segmentation model of the IOM (shown in pink), radius (R) and ulna (U) (shown in transparent blue) from the micro-CT acquisition. (c) Micro-CT-based 3D models aligned to standard-CT-based 3D bone models. (d) Separated 3D models of the IOM structures.

**Figure 8.** Thickness of IOM structures along the axial axis (left column) and the ligament fibre (right column).

**Figure 9.** Thickness variation across the longitudinal axis of the forearm. (a) Thickness at the radial attachment of the ligaments, (b) thickness at the centre of the ligament fibre, and (c) thickness at ulnar attachment. CB (yellow squares), AB (red left triangles), DOAC (orange right triangles), DOB (green stars). The corresponding linear fit is depicted by a black line, together with the fitting found by McGinley, et al. [58] , shown by a grey line.

**Tables**

| IOM Ligament | Radial Attachment (%) | | | Ulnar Attachment (%) | | | Fan-out Angle θ (°) | | |
|---|---|---|---|---|---|---|---|---|---|
| | Mean | SD | Range | Mean | SD | Range | Mean | SD | Range |
| **CB** | | | | | | | | | |
| Proximal | 65.1 | 4.2 | 59.2, 70.8 | 46.4 | 5.1 | 39.3, 54.0 | 22.5 | 6.3 | 17.9 , 34.8 |
| Distal | 51.7 | 4.0 | 44.3, 56.3 | 32.9 | 4.5 | 26.6, 40.0 | 32.6 | 5.9 | 26.8 , 43.2 |
| **AB** | | | | | | | | | |
| Proximal | 56.4 | 10.2 | 48.4, 73.5 | 37.5 | 13.0 | 24.6, 58.4 | 28.1 | 6.1 | 21.7 , 39.4 |
| Distal | 48.6 | 10.2 | 39.3, 65.6 | 28.5 | 10.2 | 19.8, 45.3 | 28.3 | 7.5 | 19.0 , 40.1 |
| **DOAC** | | | | | | | | | |
| Middle | 61.3 | 5.2 | 54.9, 69.0 | 63.1 | 2.7 | 60.4, 67.7 | -18.7 | 3.8 | -24.4 , -14.8 |
| **DOB*** | | | | | | | | | |
| Proximal | 17.5 | - | - | 19.1 | - | - | -32.8 | - | - |
| Distal | 10.0 | - | - | 11.0 | - | - | -22.9 | - | - |

*DOB was identified only 1 out of 5 specimens. Value shown corresponds to only one forearm.

**Table 1.** Attachment locations for the radial and ulnar insertions of each IOM ligament. Radial and ulnar attachments values are expressed as the relative percentage of the corresponding bone length from the distal end. The fan-out angle of each ligament fibre is given with respect to the longitudinal axis of the radius.

| IOM Ligament | 3D Method (mm) | | | Ex-Vivo (mm) | | |
|---|---|---|---|---|---|---|
| | Mean | SD | Range | Mean | SD | Range |
| **CB** | | | | | | |
| Radial Origin | 33.9 | 5.1 | 27.3 , 39.7 | 34.9 | 4.9 | 27.1 , 39.8 |
| Ulnar Insertion | 36.4 | 7.7 | 26.9 , 48.1 | 44.1 | 5.7 | 36.2 , 50.5 |
| **AB** | | | | | | |
| Radial Origin | 19.6 | 3.2 | 14.2 , 21.9 | 18.3 | 7.3 | 6.9 , 27.3 |
| Ulnar Insertion | 24.5 | 10.1 | 9.7 , 37.6 | 24.7 | 7.5 | 17.8 , 37.8 |
| **DOAC** | | | | | | |
| Radial Insertion | - | - | - | 9.2 | 2.0 | 6.5 , 12.0 |
| Ulnar Origin | - | - | - | 9.3 | 1.7 | 7.4 , 11.9 |
| **DOB**[*] | | | | | | |
| Radial Insertion | 19.6 | - | - | 18.3 | - | - |
| Ulnar Origin | 27.8 | - | - | 28.2 | - | - |

[*] DOB was identified only 1 specimen. Value corresponds to only one forearm

**Table 2.** Width of IOM structures using 3D and ex-vivo methods. The ex-vivo width of the ligament was obtained using a digital micrometre.

| IOM Ligament | 3D Method (mm) | | | CCD Laser (mm) | | |
|---|---|---|---|---|---|---|
| | Mean | SD | Range | Mean | SD | Range |
| CB | 3.52 | 1.54 | 0.85 , 6.67 | 3.44 | 1.99 | 0.12 , 7.19 |
| AB | 3.11 | 1.73 | 0.10 , 7.52 | 2.84 | 1.68 | 0.09 , 6.57 |
| DOAC | 4.65 | 1.67 | 1.26 , 7.95 | 3.79 | 1.92 | 0.62 , 6.73 |
| DOB* | 6.95 | 4.21 | 0.78 , 14.10 | 2.17 | 2.72 | 0.14 , 8.56 |

*Value shown corresponds to only forearm 4

**Table 3.** Average thickness of IOM structures

| IOM Ligament | CSA (mm²) | First Peak Force (N) | Ultimate Force (N) | Stiffness (N/mm) | Ultimate Failure Strain (%) |
|---|---|---|---|---|---|
| **CB** | | | | | |
| Forearm 1 | 130.00 | 68.81 | 98.77 | 33.09 | 3.36 |
| Forearm 2 | 149.32 | 132.60 | 196.57 | 27.71 | 3.80 |
| Forearm 3 | 105.11 | 136.80 | 188.59 | 43.48 | 3.43 |
| Forearm 4 | 134.65 | 90.09 | 186.90 | 34.02 | 6.15 |
| Forearm 5 | 161.87 | 210.70 | 268.80 | 79.33 | 5.20 |
| **Mean** | **136.19** | **127.80** | **187.93** | **43.53** | **4.40** |
| **SD** | **19.18** | **48.74** | **53.97** | **18.61** | **1.10** |
| **AB** | | | | | |
| Forearm 1 | 53.05 | 156.20 | 156.20 | 54.82 | 1.90 |
| Forearm 2 | 45.34 | 52.61 | 87.59 | 29.88 | 3.93 |
| Forearm 3 | 94.81 | 49.20 | 49.20 | 22.17 | 1.60 |
| Forearm 4 | 40.50 | 110.20 | 128.80 | 60.38 | 2.06 |
| Forearm 5 | 68.68 | 84.75 | 87.07 | 42.89 | 1.84 |
| **Mean** | **60.48** | **90.59** | **101.77** | **42.03** | **2.30** |
| **SD** | **19.64** | **39.70** | **37.08** | **14.44** | **0.80** |
| **DOAC** | | | | | |
| Forearm 1 | 38.90 | 279.60 | 327.70 | 83.93 | 2.36 |
| Forearm 2 | 46.70 | 38.58 | 46.06 | 30.32 | 1.33 |
| Forearm 3 | 34.10 | 242.40 | 259.40 | 154.40 | 1.54 |
| Forearm 4 | 21.50 | 74.04 | 80.96 | 40.04 | 1.43 |
| **Mean** | **35.30** | **158.70** | **178.50** | **77.20** | **1.70** |
| **SD** | **9.10** | **103.90** | **118.20** | **48.90** | **0.40** |
| **DOB** | | | | | |
| Forearm 5 | 50.50 | 63.40 | 69.40 | 23.57 | 1.43 |

**Table 4.** Results of tensile testing for each IOM ligament. CSA values are based on the ex-vivo measurements obtained from the CCD-Laser measurements.